\documentclass[conference]{IEEEtran}
\IEEEoverridecommandlockouts
\usepackage{cite}
\usepackage{amsmath,amssymb,amsfonts}
\usepackage{graphicx}
\usepackage{hyperref}

\usepackage{algorithm}
\usepackage{algpseudocode}

\begin{document}

\title{Enabling Differentiated QoS Degradation for Replicated Databases under Failures}

\author{
\IEEEauthorblockN{
\begin{tabular}{@{}ccc@{}}
\parbox[t]{0.31\textwidth}{\centering Belkis Djeffal} &
\parbox[t]{0.31\textwidth}{\centering Pierre Bourhis} &
\parbox[t]{0.31\textwidth}{\centering Romain Rouvoy}
\end{tabular}
}

\IEEEauthorblockA{
\begin{tabular}{@{}ccc@{}}
\parbox[t]{0.31\textwidth}{
\centering
Inria, Univ. Lille, CNRS\\
UMR 9189 CRIStAL, France\\
belkis.djeffal@inria.fr
}
&
\parbox[t]{0.31\textwidth}{
\centering
CNRS, Univ. Lille, Inria\\
UMR 9189 CRIStAL, France\\
pierre.bourhis@inria.fr
}
&
\parbox[t]{0.31\textwidth}{
\centering
Univ. Lille, Inria, CNRS\\
UMR 9189 CRIStAL, France\\
romain.rouvoy@inria.fr
}
\end{tabular}
}
}

\maketitle

\begin{abstract}  
Elasticity is commonly presented as the default response to capacity loss after failures, since replacement replicas can compensate for failed nodes and restore pre-incident service levels. Replacement capacity entails both delay and additional resource commitment, as replicas must be provisioned and synchronized before they can serve traffic. Under fixed budgets or constrained operating conditions, capacity restoration cannot be treated as the immediate recovery path. Failure handling must instead define how the service continues while capacity remains reduced. When the service exposes differentiated service levels, capacity loss cannot be handled uniformly. Degradation becomes part of the service behavior, requiring explicit control over how reduced capacity affects each class without erasing the intended differentiation.

We study differentiated QoS degradation in replicated database services with service-class-aware sessions.
We present a repair-to-target policy, implemented in PLB, a PostgreSQL JDBC middleware load balancer for service-class-aware routing. 
When fail-stop failures remove part of the available capacity, PLB updates the role assignment of healthy replicas into Premium, Mixed, and Freemium roles. 
This keeps the replica pool shared while ensuring that new session assignments continue to reflect the service class. 
We evaluate PLB under single and cascading replica failures across two deployment strategies: isolated per-class replica pools and shared, priority-agnostic routing. 
The results show that PLB improves median Premium goodput retention by 26--28 percentage points under a Premium-side fault, achieves more than \(2\times\) higher Premium goodput in the most severe cascading-failure phase, and reduces Premium p95 latency by \(18.2\%\) relative to shared round-robin.
\end{abstract}

\section{Introduction}

Fail-stop replica failures are commonly handled through failover: failed components are removed from the active set and client requests are redirected to the remaining replicas so that service can continue despite the loss of capacity~\cite{dbsurvey2000,fate2011,cjdbc2004,middler2005}. 
In replicated services, however, such failures do not end with this initial redirection step. 
They also create a post-fault regime in which the system must operate with fewer replicas, often under transient pressure from redirected demand, reconnections, and retries. 
Prior work has shown that such transient overload can substantially amplify tail latency and destabilize user-visible performance even when the service remains nominally available~\cite{tailatscale2013}.

This post-fault regime is especially challenging when the service supports multiple service classes with differentiated objectives~\cite{webtieredqos1999,slabasedcloud2011,slabasedautonomic2007}. 
After a fail-stop failure, the problem is no longer only to keep routing alive. 
The system must also decide how the remaining capacity should be allocated across service classes once the replica pool has shrunk.

Existing failover and replication mechanisms primarily focus on maintaining service continuity after failures by excluding failed replicas and redirecting requests to available replicas~\cite{dbsurvey2000,cjdbc2004,middler2005,fate2011}. 
They are, however, less explicit about how post-fault capacity should be differentiated across service classes. 
This gap matters in tiered services. 
If redirected requests are handled uniformly, high-priority and best-effort requests contend within the same smaller pool with no explicit service objective~\cite{adaptivewebadmission2004,webtieredqos1999}. 
At the other extreme, isolating service classes on separate capacity preserves separation, but under asymmetric failures it can strand healthy resources on one side of the partition while the other side comes under acute pressure~\cite{virtualpartitioning2002,slabasedautonomic2007}.

We study this problem in a replicated service built on a fixed pool of replicas serving two service classes, Premium and Freemium. 
Replica loss forces the system to continue with less capacity, without relying on immediate capacity expansion. 
The central challenge is therefore to reorganize the remaining pool so that Premium service is better preserved while the impact on Freemium remains measurable and controlled.

To address this challenge, we present a priority-aware failure-handling policy for fixed replica pools, implemented in PLB, a JDBC middleware load balancer for service-class-aware session routing. 
When fail-stop failures remove part of the available capacity, the policy uses repair-to-target to update the role assignment of healthy replicas into Premium, Mixed, and Freemium roles. 
This keeps the replica pool shared while ensuring that new session assignments continue to reflect the service class. 
In the rest of the paper, we use PLB to refer to the middleware with this failure-handling policy enabled. 
More broadly, the paper studies the tension between two post-fault choices: preserving class isolation at the risk of stranding healthy capacity, or sharing the remaining pool at the risk of losing service differentiation~\cite{virtualpartitioning2002}. 
Our objective is to show that priority-aware failure handling can navigate this tradeoff more effectively than either static isolation or priority-agnostic sharing.

Our evaluation shows that PLB changes both capacity use and service differentiation after replica failures. 
Under a Premium-side fault, PLB improves median Premium goodput retention by 26--28 percentage points and increases CPU utilization among available replicas by 18 percentage points in the Premium-heavy workload, while eliminating most of the replica imbalance. 
During cascading failures, PLB achieves over \(2\times\) higher Premium goodput in the most severe phase and reduces Premium p95 latency by 18.2\% compared with shared round-robin routing.

This paper delivers three contributions:
(1) We formalize priority-aware failure handling for replicated database services under fail-stop failures and fixed replica budgets.
(2) We present a repair-to-target failure-handling policy, implemented in PLB, that updates the role assignment of healthy replicas after faults and rejoins, converges to the configured target role counts, keeps the remaining pool available to both service classes, and gives Premium traffic preferential access to capacity.
(3) We validate the PLB implementation experimentally under single-replica and sequential two-replica failures, with comparisons to statically isolated pools and shared round-robin.

The remainder of this paper is organized as follows.
Section~II formalizes the failure-handling problem under fixed replica capacity.
Section~III presents PLB's repair-to-target policy.
Section~IV describes the experimental setup and reports the empirical evaluation.
Section~V surveys related work.
Section~VI discusses threats to validity, and Section~VII concludes.
\section{Problem Statement}\label{sec:problem}
We consider priority-aware failure handling in a replicated database
service with a fixed replica budget. A fail-stop failure reduces the
active set from \(K\) replicas to \(K^+<K\), and replacement capacity
is not assumed to become immediately available. 
The problem is therefore to reorganize and share the surviving
capacity across service classes so that Premium traffic receives
preferential access to capacity, while the resulting degradation
of Freemium traffic remains explicit and measurable.

\subsection{Failure Model}

Let \(\mathcal{R}=\{r_1,\ldots,r_K\}\) denote the initial set of replicas. 
All replicas store the same logical database and can serve sessions independently. 
The replica budget \(K\) is fixed during the failure-handling period: failed replicas are not replaced by immediately usable new capacity.

Clients interact with the service through sessions. A session \(s\) is a sequence of queries issued over a persistent database connection and is tagged by the application with \(\operatorname{class}(s)\in\{P,F\}\). Premium (\(P\)) denotes the higher-priority class to which PLB gives preferential access to surviving capacity, whereas Freemium (\(F\)) denotes the best-effort class that may experience greater degradation when capacity is reduced. For example, these classes may represent customer-facing analytical sessions and delay-tolerant background analytics, respectively. The labels specify the priority ordering enforced by PLB.

We deliberately adopt session-level, non-migratory routing: PLB selects the target replica when the JDBC connection is established, and the session remains on that replica for its lifetime. Live-session migration constitutes a separate mechanism outside the scope of this work.

At time \(t\), each replica has a health state
\[
h(r,t) \in \{\textsc{Healthy},\textsc{Down}\}.
\]
The survivor set is
\[
\mathcal{R}^+(t)=\{r \in \mathcal{R} \mid h(r,t)=\textsc{Healthy}\},
\]
with \(K^+(t)=|\mathcal{R}^+(t)|\).

A fail-stop failure is represented as a sequence of events
\[
I = \langle e_1,\ldots,e_m\rangle,
\]
where each event \(e_i = (t_i,r_i,a_i)\) occurs at time \(t_i\), affects replica \(r_i\), and has action \(a_i \in \{\textsc{Down},\textsc{Rejoin}\}\). 
A \textsc{Down} event makes a replica unavailable. 
A \textsc{Rejoin} event marks the point at which a previously failed replica becomes reachable again and is returned to the active replica set. 
A single-fault episode contains one down event and, optionally, one rejoin event. 
A cascading failure episode involves multiple \textsc{Down} events before the system fully returns to its pre-fault state. If several events have the same timestamp, they appear in \(I\) in the order in which PLB observes them.

The event sequence partitions execution into observation intervals. 
Before the first down event, the system operates with the full replica set \(K\). 
Between down and rejoin events, the system operates with \(K^+(t)<K\). 
In cascading failures, several fault intervals may occur, corresponding to different values of \(K^+(t)\).

\subsection{Failure-Handling Strategy}

A failure-handling strategy determines how surviving capacity is organized across service classes and how newly arriving sessions are assigned. Let \(\Sigma(t)\) denote the routing state observed by PLB at time \(t\), consisting of the current role assignment and the per-class active-session counts of the healthy replicas. For an arriving session \(s\), an online strategy \(\Gamma\) selects a target replica:
\[
\Gamma\bigl(\operatorname{class}(s),\Sigma(t),\mathcal{R}^+(t)\bigr)
\in \mathcal{R}^+(t).
\]
At time \(t\), these arguments reflect only the events observed up to \(t\), not future events.
A priority-agnostic strategy may route all sessions uniformly over \(\mathcal{R}^+(t)\), preserving sharing but losing service differentiation. 
A static per-class partition may preserve isolation between Premium and Freemium sessions, but it can leave healthy replicas unusable for the class under pressure after an asymmetric failure. 
We seek a strategy that combines both properties: sharing the surviving replica set while preserving differentiated service behavior.

\subsection{Evaluation Criteria}

We characterize behavior after failures using per-class progress and tail-latency criteria. 
For each class \(c \in \{P,F\}\), let \(L_c^{base}\) denote a tail-latency statistic measured during the pre-fault reference interval, and let \(L_c^{I}\) denote the same statistic measured while one or more replicas are down. 
Latency inflation is
\[
\mathrm{Inflation}_c = \frac{L_c^{I}}{L_c^{base}}.
\]
Similarly, let \(G_c^{base}\) denote pre-fault goodput and \(G_c^{I}\) denote goodput while one or more replicas are down. 
Goodput retention is
\[
\mathrm{Retention}_c = \frac{G_c^{I}}{G_c^{base}}.
\]
Lower latency inflation and higher goodput retention indicate that a class remains closer to its pre-fault service level. 
In the empirical evaluation, we instantiate \(L_c\) with p95 latency and \(G_c\) with completed-query goodput. 
We additionally report observed session lifetime as an indicator of backlog pressure for finite-lifetime sessions.

\subsection{Design Goals}

We characterize the expected behavior of a priority-aware failure-handling strategy through three goals.

\textbf{G1: Preserve high-priority service under capacity loss.}
While replicas are unavailable, Premium service should remain closer to its pre-fault behavior than under priority-agnostic failover. 
We evaluate this through lower Premium latency inflation, higher Premium goodput retention, and shorter observed session lifetime.

\textbf{G2: Avoid stranded capacity and adapt to failure shape.}
When failures are asymmetric, healthy replicas should not remain unusable only because they were associated with another service class before the fault. 
The strategy should reorganize the survivor set according to the current value of \(K^+(t)\), including when failures occur sequentially.

\textbf{G3: Make low-priority impact measurable and controlled.}
Protecting Premium shifts part of the capacity-loss cost to Freemium traffic. 
During severe failures, Freemium sessions may receive lower goodput, higher latency, or longer observed session lifetime.
The goal is not to hide this cost, but to keep it observable through service-level metrics and assess whether Freemium continues to make progress.

\subsection{Problem Formulation}

Given a fixed set \(\mathcal{R}\) of \(K\) replicas, a stream of sessions tagged with service classes \(P\) and \(F\), and a failure sequence \(I\) that changes the active survivor set from \(K\) replicas to \(K^+(t)<K\), we seek a failure-handling strategy \(\Gamma\) that operates only on the surviving replicas and satisfies G1--G3. 
The strategy should limit high-priority QoS degradation when capacity is missing, avoid stranding healthy capacity behind static class boundaries, including during cascading failures, and make low-priority impacts measurable and controlled.
\section{Differentiated QoS Control under Replica Failures}\label{sec:approach}

\subsection{Overview}
We now describe how PLB reacts once a replica fault reduces the active replica set. The service must continue with fewer healthy replicas, and the remaining capacity must be redistributed across service classes.

PLB is implemented as a JDBC middleware that assigns each newly admitted session to a replica according to its service class and the observed replica load. It organizes healthy replicas into three logical roles: Premium and Freemium roles provide class-preferred capacity, while the Mixed role provides shared capacity. During normal load variations, replicas may be temporarily borrowed across roles and returned when pressure subsides.

After a replica fault, keeping the surviving role composition unchanged would make the post-fault layout depend on which replica failed, whereas collapsing the roles into a shared pool would remove service differentiation. PLB therefore computes a target Premium/Mixed/Freemium role vector for the current survivor set and repairs the role assignment toward that target. The repair affects subsequent admissions, while running sessions remain pinned.

\begin{figure}[t]
  \centering
  \includegraphics[width=0.82\columnwidth]{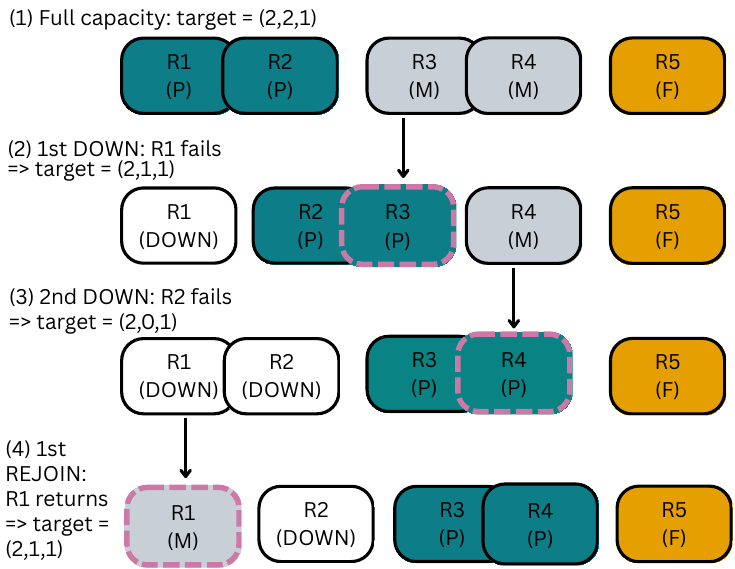}
  \caption{Example repair-to-target sequence for a five-replica pool with reference split \((K_P^0,K_M^0,K_F^0)=(2,2,1)\). Teal, gray, orange, and white denote Premium, Mixed, Freemium, and \textsc{Down}; pink dashed outlines mark role changes.}
  \label{fig:plb-repair-sequence}
\end{figure}

Figure~\ref{fig:plb-repair-sequence} illustrates this repair-to-target process.

\subsection{Replica State after Faults and Rejoins}
To apply repair-to-target, the policy distinguishes the full-capacity reference split from the current post-fault role assignment. 
The reference split is the configured PLB allocation, denoted by
\[
(K_P^0,K_M^0,K_F^0),
\qquad
K_0 = K_P^0+K_M^0+K_F^0 .
\]
It records the role balance of the original deployment and is not modified by individual faults or borrowing decisions.

While replicas are unavailable, the policy operates on the remaining healthy replicas. 
At a given point, let \(\mathcal{R}^+\) denote this survivor set, with \(K^+=|\mathcal{R}^+|\). 
Before repair, transient borrowed states are canonicalized into three mutually exclusive role sets:
\begin{align}
\mathcal{R}^+ &=
\mathcal{R}_P \cup \mathcal{R}_M \cup \mathcal{R}_F, \\
\mathcal{R}_P \cap \mathcal{R}_M &= \emptyset,\qquad
\mathcal{R}_P \cap \mathcal{R}_F = \emptyset,\qquad
\mathcal{R}_M \cap \mathcal{R}_F = \emptyset .
\end{align}
Here, \(\mathcal{R}_P\), \(\mathcal{R}_M\), and \(\mathcal{R}_F\) denote the healthy replicas whose current roles are Premium, Mixed, and Freemium. 
Their sizes are
\begin{align}
K_P &= |\mathcal{R}_P|,\qquad
K_M = |\mathcal{R}_M|,\qquad
K_F = |\mathcal{R}_F|, \\
K^+ &= K_P + K_M + K_F .
\end{align}

Canonicalization removes transient borrowed states before the target is applied. 
A healthy replica carrying sessions from both classes is placed in \(\mathcal{R}_M\); a replica carrying only Premium sessions is placed in \(\mathcal{R}_P\); and a replica carrying only Freemium sessions is placed in \(\mathcal{R}_F\). 
Borrowed replicas with no active borrower-class sessions are returned to one of these three roles. 
After this step, each healthy replica belongs to exactly one role set.

Faults and rejoins only change the survivor set. 
When a replica becomes unavailable, it is removed from \(\mathcal{R}^+\); upon rejoining, it is added back and treated initially as shared capacity before repair is applied. 
The same state representation is therefore used after a single fault, after cascading faults, and during partial recovery.

\subsection{Repair-to-Target Policy}

Given the canonical role partition, the policy computes the role sizes that should be maintained with the current active set. 
We call this vector the target role vector:
\[
\mathbf{K}^{I}(K^+) =
(K_P^{I}, K_M^{I}, K_F^{I}),
\]
where \(K_P^{I}\), \(K_M^{I}\), and \(K_F^{I}\) denote the target numbers of Premium, Mixed, and Freemium replicas for the current failure sequence \(I\). 
The superscript \(I\) indicates that the target is recomputed as the active replica set changes during the failure sequence.

For small active sets, the target is defined explicitly to keep the repair policy well-defined under severe cascading loss:
\[
\mathbf{K}^{I}(1)=(0,1,0),\quad
\mathbf{K}^{I}(2)=(1,0,1),
\]
\[
\mathbf{K}^{I}(3)=(2,0,1).
\]
These boundary cases define the policy when too few replicas remain for proportional scaling to determine a meaningful integer role allocation. With one survivor, the Mixed role keeps the remaining capacity accessible to both classes; with two, the target preserves one role per class; with three, the additional role is assigned to Premium. For \(K^+\geq4\), the target is obtained by proportionally scaling the full-capacity reference allocation and adjusting it to satisfy the replica budget and role floors. The evaluation does not rely on the one- and two-replica cases.

For larger active sets, the target role vector follows the full-capacity reference allocation. 
Let
\[
(K_P^0,K_M^0,K_F^0)
\]
denote the configured role sizes under the full replica set, with
\[
K_0 = K_P^0 + K_M^0 + K_F^0 = K.
\]
For \(K^+\geq 4\), the policy first obtains tentative Premium and Freemium targets by nearest-integer proportional scaling:
\[
\widetilde K_P =
\max\left(1,\operatorname{round}\left(K^+\frac{K_P^0}{K_0}\right)\right),
\]
\[
\widetilde K_F =
\begin{cases}
\max\left(1,\operatorname{round}\left(K^+\frac{K_F^0}{K_0}\right)\right) & \text{if } K_F^0>0,\\
0 & \text{otherwise}.
\end{cases}
\]

These tentative values are then adjusted to satisfy the active-set budget and the role floors. 
The resulting target role vector
\[
\mathbf{K}^{I}(K^+)=(K_P^I,K_M^I,K_F^I)
\]
satisfies:
\[
K_P^I+K_M^I+K_F^I=K^+,
\]
with the Mixed role receiving the residual capacity:
\[
K_M^I = K^+ - K_P^I - K_F^I .
\]

Algorithm~\ref{alg:repair-to-target} summarizes the repair step. 
The procedure canonicalizes the current role state, computes the target role vector, and updates role membership until the current role sizes match the target. 
When a role is below its target, the policy considers donors from \(\mathcal{R}_M\) first, since Mixed is the flexible role. 
Moving capacity across the Premium/Freemium boundary is allowed only when the corresponding floor remains protected. 
Among eligible donors, the least-loaded replica is selected.
A Mixed replica can change role while serving both classes; existing sessions continue normally, and only new sessions follow the new role.

\begin{algorithm}[!t]
\caption{Repair-to-target after an active-set change}
\label{alg:repair-to-target}
\small
\begin{algorithmic}[1]
\Require Current role sets \(\mathcal{R}_P,\mathcal{R}_M,\mathcal{R}_F\); reference split \((K_P^0,K_M^0,K_F^0)\)
\Ensure Repaired role sets \(\mathcal{R}_P,\mathcal{R}_M,\mathcal{R}_F\)

\State Canonicalize borrowed states into \(\mathcal{R}_P,\mathcal{R}_M,\mathcal{R}_F\)
\State \(\mathcal{R}^+ \gets \mathcal{R}_P \cup \mathcal{R}_M \cup \mathcal{R}_F\)
\State \(K^+ \gets |\mathcal{R}^+|\)
\State \((K_P^I,K_M^I,K_F^I) \gets \mathbf{K}^{I}(K^+)\)

\While{\((|\mathcal{R}_P|,|\mathcal{R}_M|,|\mathcal{R}_F|)
      \neq (K_P^I,K_M^I,K_F^I)\)}
    \If{\(|\mathcal{R}_P| < K_P^I\)}
        \State Move the least-loaded eligible donor to \(\mathcal{R}_P\)
    \ElsIf{\(|\mathcal{R}_F| < K_F^I\)}
        \State Move the least-loaded eligible donor to \(\mathcal{R}_F\)
    \ElsIf{\(|\mathcal{R}_P| > K_P^I\)}
        \State Move the least-loaded eligible replica from \(\mathcal{R}_P\) to \(\mathcal{R}_M\)
    \ElsIf{\(|\mathcal{R}_F| > K_F^I\)}
        \State Move the least-loaded eligible replica from \(\mathcal{R}_F\) to \(\mathcal{R}_M\)
    \Else
        \State \textbf{break}
    \EndIf
\EndWhile

\State \Return \(\mathcal{R}_P,\mathcal{R}_M,\mathcal{R}_F\)
\end{algorithmic}
\end{algorithm}
\normalsize

The output of repair-to-target is the role partition used by subsequent admissions. 
The policy revises the role structure while session assignment remains load-aware. 
Since \(\mathbf{K}^{I}(K^+)\) is recomputed after every change to \(\mathcal{R}^+\), the same procedure covers faults affecting any role, cascading faults, and partial recovery.
\section{Empirical Evaluation}\label{sec:evaluation}
We evaluate PLB under fail-stop replica failures in a fixed five-replica deployment. Each run captures the system before the fault, while one or more replicas are unavailable, and after replicas rejoin. This structure distinguishes normal, post-fault, and recovery behavior.

Our evaluation contrasts two deployment choices commonly available to a provider. Dedicated per-class clusters enforce service differentiation through isolation: each class is bound to its own subset of replicas. This avoids cross-class interference, but makes the allocation rigid when failures affect only one side of the partition. Shared round-robin routing removes this static boundary and allows all sessions to use the remaining replicas, but it does not encode the service class in the failover decision. PLB occupies the space between these two choices by keeping the pool shared while adapting the role layout and routing policy after failures.
\subsection{Research Questions}

We organize the evaluation around three questions, aligned with the design goals introduced in Section~\ref{sec:problem}. 
They test whether PLB avoids stranded capacity, preserves service differentiation, and adapts as replicas fail and rejoin.

\textbf{\phantomsection\label{rq:static-isolation}RQ1.}
How does static per-class isolation shape the use of remaining capacity under class-local replica failures?

\textbf{\phantomsection\label{rq:shared-degraded-pool}RQ2.}
How does PLB preserve service differentiation in a shared post-fault replica pool?

\textbf{\phantomsection\label{rq:adaptivity}RQ3.}
How does PLB adapt as failures cascade and replicas rejoin?

\subsection{Experimental Methodology}
\label{sec:experimental-setup}

\paragraph{Deployment and workload}
We run the experiments on Grid'5000~\cite{grid5000}, using a physically distributed deployment with five PostgreSQL replica instances and one workload generator. 
The experiments use homogeneous machines from the \texttt{gros} cluster at the Nancy site, each equipped with an 18-core Intel Xeon Gold 5220 CPU and 96~GiB of RAM. 
All nodes run Ubuntu 22.04 LTS, PostgreSQL 14, and OpenJDK 21. 
The replicas store identical copies of the same database and can serve sessions independently. 
The workload generator connects via the PLB middleware, which assigns each admitted session to a replica based on the selected strategy. 
Throughout each run, the replica budget remains fixed: when a replica fails, it is excluded from the active set, and no replacement capacity is provisioned.

The workload is generated with BenchBase using TPC-H~\cite{tpch,DifallahPCC13}. 
We extend BenchBase to support service-class tags, Poisson arrivals, and finite session lifetimes. 
Each session is labeled as Premium or Freemium, opens a JDBC connection, and remains assigned to the selected replica for its lifetime. 
We use seeded workload generation to make runs reproducible and, where applicable, to align arrival structure across compared strategies. 
Before collecting measurements, we execute a warmup phase to initialize the database and benchmark state; the reported measurements are collected during the subsequent monitored run.

\paragraph{Configurations}
We vary the total client load \(T\) and the composition of the workload. 
A mix \(Px{:}Fy\) denotes a workload in which \(x\%\) of clients are Premium and \(y\%\) are Freemium. 
The evaluation uses three mixes: \(P25{:}F75\) for Freemium-heavy demand, \(P50{:}F50\) for balanced demand, and \(P75{:}F25\) for Premium-heavy demand. 
For a given configuration, compared strategies use the same client load, workload mix, and fault schedule, so that observed differences come from routing and fault-time role reassignment rather than from changes in offered load.

\paragraph{Fault injection and phase windows}
We evaluate fail-stop replica faults. 
A down event transitions a replica to the \textsc{Down} state, excluding it from new session assignments. 
A later rejoin event returns the replica to the active set. 
We use two fault patterns. 
Single-fault runs contain one down event followed by a rejoin. 
Cascading-fault runs contain two sequential down events before the rejoin phases. 
These patterns allow us to evaluate both an isolated capacity loss and a more severe failure episode in which the active replica set is reduced in steps.

Each run is divided into phase windows around the down and rejoin events. 
The interval immediately preceding the first fault is used as the reference interval. 
We exclude the initial ramp-up because client arrivals are still building up during that period, and including it would make the baseline less representative of steady pre-fault behavior. 
Fault phases correspond to intervals in which one or more replicas are down. 
Rejoin phases capture the period after failed replicas return to the active set, where capacity has been restored, but service-level metrics may still reflect work admitted earlier. 
The affected replica role and the exact event times are provided in the corresponding research question, as they differ across the evaluated fault patterns.

\paragraph{Metrics, aggregation, and artifacts}
We collect per-class goodput, 95th-percentile latency (p95), and observed session lifetime. 
Goodput is the number of successfully completed benchmark queries per second within a phase window, computed separately for Premium and Freemium sessions. 
Failed, timed-out, and incomplete queries are excluded. 
The p95 latency captures tail service quality, which is important because overload and fault-induced pressure often appear first in the tail of the latency distribution~\cite{tailatscale2013}. 
Observed session lifetime measures how long sessions remain active, including delay caused by post-fault execution, and serves as an indicator of backlog pressure. 
For the resource-use analysis in RQ1, we also collect per-replica CPU utilization.

For each configuration, we repeat the experiment 6 times and aggregate results across runs using medians. 
Phase-level results are computed within each phase window. 
The experiment scripts, middleware implementation, and modified BenchBase are available as an artifact.\footnote{\url{https://zenodo.org/records/20121219}}

\subsection{Baselines}

We compare PLB against two deployment choices. 
For a replica budget \(K\), we consider:
\begin{enumerate}
    \item \textbf{Dedicated per-class partitions (\(r\)\textsf{RR}+\(f\)\textsf{RR}, \(r+f=K\)).}
    The \(K\) replicas are statically partitioned between service classes. 
    The left-hand term denotes the number of replicas assigned to Premium traffic, and the right-hand term denotes the number of replicas assigned to Freemium traffic. 
    Within each partition, sessions are routed using round-robin. 
    Under a class-local failure, the affected partition continues with fewer replicas, while the other partition remains isolated.

    \item \textbf{Shared, priority-agnostic pool (\(K\)\textsf{RR}).}
    All sessions share the same set of \(K\) replicas, and connections are distributed using round-robin over the currently available replicas. 
    When a replica fails, it is removed from the active set, and sessions continue to share the remaining replicas without regard to service class.
\end{enumerate}

PLB occupies the middle ground between these baselines: it starts from the same fixed replica budget, but repairs the post-fault role layout after failures and makes routing decisions service-class aware. 
Unless otherwise stated, we instantiate the dedicated baseline as \textsf{3RR+2RR} and the shared baseline as \textsf{5RR}.

\subsection{RQ1: Class-Local Failures in Statically Isolated Pools}\label{sec:rq1-static-isolation}

We first evaluate a class-local fail-stop fault, in which the loss of one replica reduces the capacity available to Premium while the Freemium replicas remain healthy.

This setting captures a common failure mode of statically isolated deployments: capacity may still exist in the system, but it cannot be reused across the class boundary. 
We compare PLB with a static \textsf{3RR+2RR} configuration under the same five-replica budget. 
In \textsf{3RR+2RR}, three replicas are dedicated to Premium sessions and two replicas to Freemium sessions, with round-robin routing within each group.

We use \(T=1000\) clients generated with the Poisson-arrival workload described in Section~\ref{sec:experimental-setup}, and consider three Premium:Freemium mixes: \(25{:}75\), \(50{:}50\), and \(75{:}25\). 
A fail-stop fault is injected at \(t=120\)s on a replica assigned to Premium traffic; the replica becomes available again at \(t=210\)s. 
During the fault interval, Premium has fewer dedicated replicas, while Freemium-side replicas remain unavailable to Premium under the static partition.

We evaluate the fault interval using one service-level metric and two resource-level metrics. 
Premium goodput retention measures how much useful Premium progress is preserved while the Premium partition has lost one replica:
\[
  \mathrm{Retention}_{P}
  =
  \frac{
    \mathrm{Goodput}_{P}^{[120,210)}
  }{
    \mathrm{Goodput}_{P}^{[90,120)}
  }.
\]
The denominator uses the 30 seconds immediately preceding the fault. 
This choice excludes the initial Poisson-arrival ramp-up and normalizes fault-window goodput by the Premium goodput measured immediately before capacity is reduced. 
A value of one indicates that Premium preserves its pre-fault goodput.

For resource usage, let \(R^{+}\) be the set of available replicas during the fault interval and let \(C\) be the number of CPU cores per replica. 
For each replica \(r \in R^{+}\), we compute its mean CPU fraction over a window \(W\) as
\[
  u_r(W)
  =
  \frac{1}{|W|}
  \sum_{t \in W}
  \frac{\mathrm{cpu}_{r,t}^{user} + \mathrm{cpu}_{r,t}^{sys}}{100 \cdot C}.
\]
The mean CPU fraction over the available replicas is then
\[
  \bar{u}(W) = \frac{1}{|R^{+}|}\sum_{r \in R^{+}} u_r(W),
\]
and the replica imbalance is measured with the coefficient of variation
\[
  \mathrm{CV}(W)
  =
  \frac{\sigma(\{u_r(W)\}_{r \in R^{+}})}
       {\mu(\{u_r(W)\}_{r \in R^{+}})}.
\]
Thus, \(\bar{u}\) captures how much of the remaining capacity is used, while \(\mathrm{CV}\) captures how evenly this load is distributed across available replicas.

\begin{figure}[t]
  \centering
  \includegraphics[width=\columnwidth]{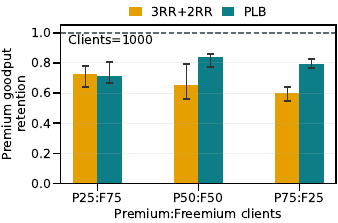}
  \caption{Premium goodput retention under a Premium-side fail-stop fault at \(T=1000\). Bars show medians; error bars show interquartile ranges.}
  \label{fig:rq1-goodput}
\end{figure}

\paragraph{Premium goodput retention}
Figure~\ref{fig:rq1-goodput} reports the effect of the Premium-side fault on completed Premium queries. 
Across all mixes, PLB retains more pre-fault Premium goodput than the dedicated priority-cluster baseline. 
The difference is modest when Premium is a minority, but becomes substantial as Premium demand increases: the median Premium goodput retention under PLB exceeds that of the dedicated baseline by 26 percentage points at \(P50{:}F50\) and by 28 percentage points at \(P75{:}F25\).
This shows that static isolation is most costly when the class affected by the fault is also the class placing pressure on the system.

\begin{figure}[t]
  \centering
  \includegraphics[width=\columnwidth]{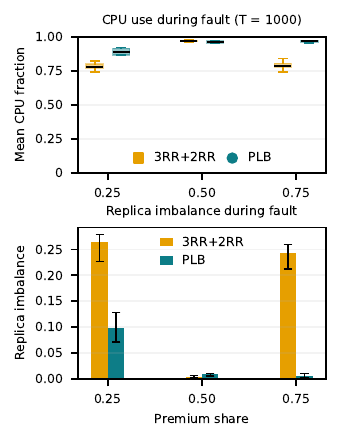}
  \caption{CPU fraction and replica imbalance among available replicas during the fault interval at \(T=1000\).}
  \label{fig:rq1-cpu}
\end{figure}

\paragraph{Capacity use during the fault}
Figure~\ref{fig:rq1-cpu} reports how the remaining replica pool is used during the fault interval. 
The top panel shows that the static split becomes inefficient when the post-fault workload no longer matches the fixed partition. 
This is most visible when Premium dominates (\(P75{:}F25\)): the dedicated priority-cluster baseline uses only 0.79 of the available CPU capacity, whereas PLB reaches 0.97. 
Since the fault removes capacity from the Premium side, the static baseline keeps Premium confined to a smaller partition even though other replicas remain healthy. 
Conversely, when Freemium dominates (\(P25{:}F75\)), Premium demand is too small to fully use the remaining Premium-side capacity, while most sessions are Freemium. 
PLB avoids this mismatch and raises the mean CPU fraction from 0.78 to 0.89.

The bottom panel shows that this mismatch also appears as a replica-level imbalance. 
In the two skewed mixes, the dedicated baseline has CV values around 0.24--0.26, meaning that some available replicas carry substantially more load than others. 
By contrast, PLB reduces CV to 0.10 at \(P25{:}F75\) and nearly zero at \(P75{:}F25\). 
The balanced mix follows the same pattern: when the workload is close to the static split, both configurations keep the available replicas highly utilized and nearly balanced. 
Taken together, these measurements answer RQ1 by showing that static isolation remains rigid after a class-local fault. 
Capacity continues to follow the pre-fault partition, whereas PLB adapts the role layout to the workload observed after the fault, addressing the stranded-capacity goal G2.

\subsection{RQ2: Preserving Service Differentiation in a Shared Post-Fault Pool}\label{sec:rq2-shared-postfault-pool}

Having examined static isolation, we next consider the shared-pool alternative. 
A shared pool avoids stranding healthy replicas behind fixed class boundaries, but sharing alone does not preserve service differentiation. 
If the routing policy remains priority-agnostic, Premium and Freemium sessions contend for the same smaller pool with no service-class objective. 
RQ2, therefore, asks how PLB changes the behavior of a shared pool after failures and how the resulting cost is distributed across service classes.

\begin{figure*}[t]
  \centering
  \includegraphics[width=\textwidth]{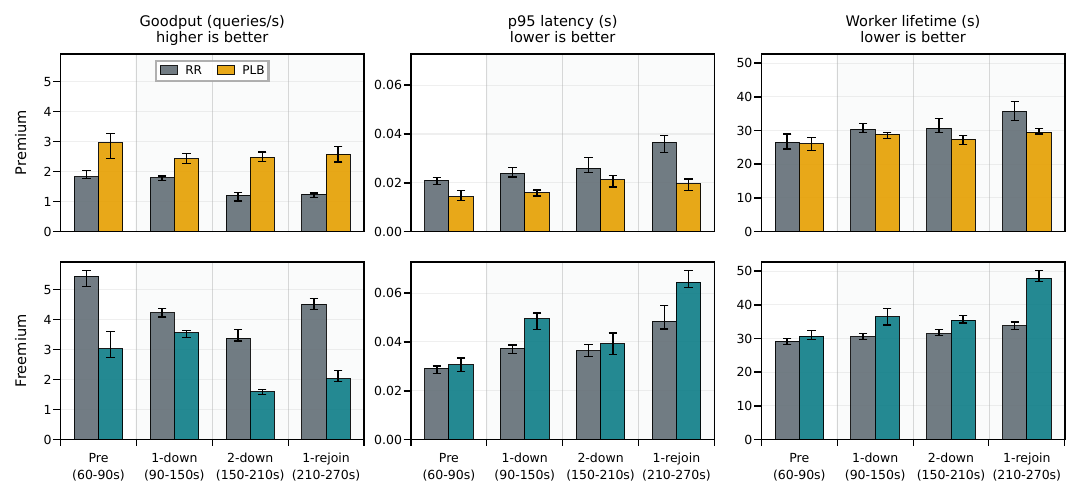}
  \caption{Phase-summary behavior during a cascading failure at \(T=1000\), \(P25{:}F75\).}
  \label{fig:rq2-phase}
\end{figure*}

We compare PLB with \textsf{5RR} under a cascading failure in which two replicas fail sequentially before recovery begins. 
Each run is summarized over the phase intervals shown on the x-axis of Fig.~\ref{fig:rq2-phase}: a pre-fault reference interval, one replica down, two replicas down, and partial recovery after one replica rejoins. 
We focus first on the Freemium-heavy workload \(T=1000\), \(P25{:}F75\). 
This setting stresses service differentiation because Premium is the minority class: under priority-agnostic sharing, the shared pool naturally follows the Freemium-dominated arrival mix unless capacity is explicitly redirected.

Figure~\ref{fig:rq2-phase} reports phase-level goodput, p95 latency, and observed session lifetime, using the aggregation described in Section~\ref{sec:experimental-setup}. 
Together, these metrics distinguish whether priority preservation is only a latency effect or also affects progress and backlog pressure after faults occur. 
Each bar summarizes one phase, using the median across repeated runs; error bars show the interquartile range.

The figure shows a consistent trade-off between protection and cost. 
For Premium, PLB preserves more useful work than \textsf{5RR}: goodput remains higher, while p95 latency and observed session lifetime remain lower across the fault phases. 
In the two-replica-down interval, where pressure is strongest, Premium goodput under PLB is more than twice that of \textsf{5RR}, while Premium p95 latency is reduced by \(18.2\%\). 
Thus, Premium sessions maintain progress and experience lower tail latency while the system operates with fewer replicas.

Freemium exhibits the complementary effect. 
It completes less work and sees higher p95 latency and observed session lifetime, reflecting the cost of preserving Premium under a fixed replica budget. 
Since cascading failures remove Premium-side capacity, maintaining Premium service requires allocating a larger share of the remaining pool to Premium sessions. 
PLB, therefore, does not eliminate the failure cost; it makes the service-class trade-off visible in the lower-priority class.

\begin{figure}[t]
  \centering
  \includegraphics[width=\columnwidth]{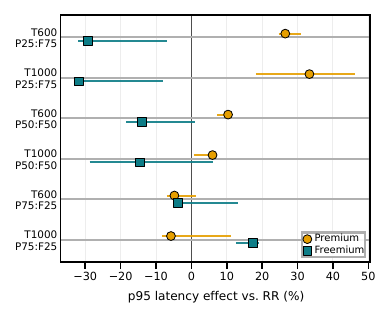}
  \caption{p95 latency effect relative to \textsf{5RR} across loads and workload mixes. Positive values indicate lower p95 latency under PLB.}
  \label{fig:rq2-forest}
\end{figure}

Figure~\ref{fig:rq2-forest} summarizes this effect across the workload grid using a common sign convention for both service classes. 
Positive values indicate that PLB reduces p95 latency relative to \textsf{5RR}, whereas negative values indicate higher p95 latency. 
With this convention, the separation between the Premium and Freemium blocks visualizes the differentiated QoS effect induced by PLB: Premium moves mostly toward improvement, while Freemium moves toward degradation. 
The distance from zero captures the magnitude of the effect, showing that Premium improvements and Freemium slowdowns are of comparable order. 
The priority shift is most visible when Premium is not already the dominant class: across the \(P25{:}F75\) and \(P50{:}F50\) mixes, PLB reduces Premium p95 latency by \(6.0\%\) to \(33.4\%\), while Freemium slowdowns range from \(14.0\%\) to \(31.8\%\). 
In Premium-dominant workloads, the gap to \textsf{5RR} naturally narrows because Premium sessions already dominate the shared pool. 
Under a fixed replica budget, PLB cannot create additional capacity; it has less low-priority demand from which to shift resources toward Premium.

These results answer RQ2 by separating the effects of sharing from those of priority-aware control. 
Shared round-robin avoids the rigidity of isolated pools, but it treats all sessions uniformly after capacity loss. 
PLB changes the allocation of the same replica budget: Premium keeps higher progress and lower tail latency, while Freemium absorbs a larger, measured share of the failure cost, matching the protection-cost trade-off captured by G1 and G3.

\subsection{RQ3: Adaptation Across Cascading Failures and Rejoins}
\label{sec:rq3-adaptivity}

We finally evaluate PLB under a sequential two-fault scenario. 
This setting exercises PLB's repair-to-target policy as the active replica set shrinks and later expands. 
The fault sequence first removes one Premium replica, then a second, before the failed replicas rejoin sequentially. 
The run therefore moves through full capacity, one-replica loss, two-replica loss, and the first and second rejoin phases.

\begin{figure*}[t]
  \centering
  \includegraphics[width=\textwidth]{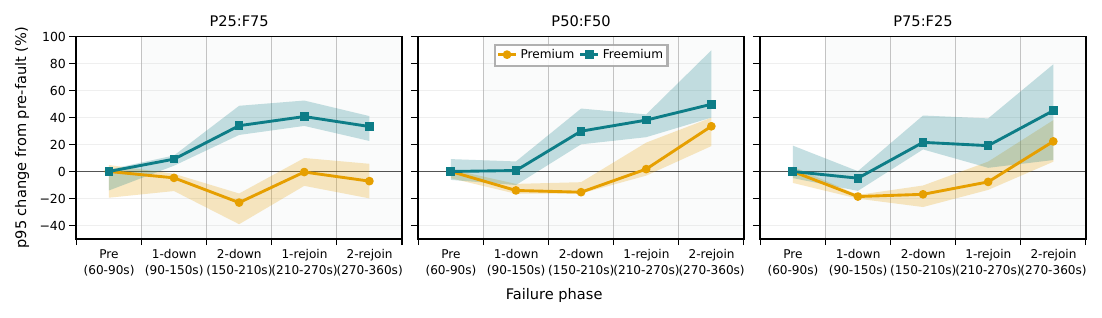}
  \caption{PLB response during a sequential Premium-side two-fault scenario at \(T=600\).}
  \label{fig:rq3-p95-change}
\end{figure*}

A Premium-side cascading failure is a stringent test for PLB because it reduces capacity in the replica subset associated with the protected service class. 
We run the experiment at a client load \(T=600\) and include three workload mixes: \(P25{:}F75\), \(P50{:}F50\), and \(P75{:}F25\), corresponding to Freemium-heavy, balanced, and Premium-heavy demand. 
Figure~\ref{fig:rq3-p95-change} reports the percentage change in 95th-percentile latency (p95) relative to the interval immediately preceding the first failure. 
Each class is normalized by its own reference p95. 
The horizontal line at \(0\%\) denotes this reference level; positive values indicate p95 degradation, while negative values indicate that p95 remains below the reference.

During the capacity-loss phases, Premium remains below its reference p95 across the three workload mixes. 
In the one-down phase, Premium p95 changes by \(-18.6\%\) to \(-4.7\%\), and in the two-down phase by \(-23.0\%\) to \(-15.3\%\). 
This shows that PLB continues to protect Premium tail latency even when the failure affects Premium-side capacity. 
This result is consistent with PLB's repair-to-target policy: after each change in the active replica set, PLB reorganizes the remaining replicas so that new assignments continue to favor high-priority service.

The lower-priority class reflects the cost of this choice. 
In the two-down phase, Freemium p95 increases by \(21.6\%\) to \(33.9\%\). 
Under a fixed replica budget, this increase is the trade-off for preserving Premium tail latency while two replicas are unavailable. 
The separation between the two classes is largest in this phase: the p95 change in Freemium-minus-Premium ranges from \(38.5\) to \(57.0\) percentage points across the three mixes. 
Thus, PLB does not remove the failure cost; it changes how that cost is distributed across service classes.

The rejoin phases introduce an important qualification. 
Restoring replica availability does not immediately restore the service-level latency distribution observed before the failure. 
Freemium remains above its reference p95 after the replicas return, and in the final phase, Premium also increases for the balanced and Premium-heavy mixes. 
This suggests a recovery-tail effect, where work delayed during the two-down interval continues to influence later completions after replicas become available again. 
PLB updates replica roles after faults and rejoins; however, it does not explicitly manage draining the accumulated work while capacity was missing.

The sequential scenario confirms that PLB reacts to changes in the active replica set rather than to a single fixed fault pattern. 
Each \textsc{Down} or \textsc{Rejoin} event triggers a new repair target, and the resulting assignments preserve Premium tail latency during the capacity-loss phases. 
The post-rejoin phases also reveal the limit of the current design: repairing replica roles is faster than draining delayed work, so service-level recovery can lag behind replica availability.
\section{Related Work}\label{sec:related}

Our setting lies at the intersection of failure handling, replicated-database middleware, and priority-aware control under resource pressure. 
Prior work addresses these dimensions mostly in isolation. 
It can expose failures, redirect traffic after faults, or prioritize requests under load, but it does not define a priority-aware failure-handling policy for a fixed replica pool after capacity loss.

A first line of work focuses on exposing failures and validating recovery behavior in distributed systems. 
Jepsen~\cite{jepsen} popularized black-box fault injection combined with history checking. 
FATE and DESTINI~\cite{fate2011} introduced systematic recovery testing for cloud systems. 
FaultSee~\cite{faultsee2020} and Legolas~\cite{legolas2024} improved the reproducibility and effectiveness of fault-injection campaigns, including the exposure of subtle partial failures. 
Recent work has also considered database-facing resilience testing, for example, through Filibuster-style fault injection around unreliable databases~\cite{filibuster2024}, while LazyFS~\cite{lazyfs2024} studies storage-level data-loss bugs triggered by injected faults. 
These efforts are valuable for constructing realistic failure scenarios and evaluating recovery behavior. 
However, they focus on testing systems under fault conditions, not on controlling how a replicated database service should allocate its remaining capacity after a fail-stop fault.

A second line of work concerns replicated-database middleware, clustering layers, and failover mechanisms. 
C-JDBC~\cite{cjdbc2004} and MIDDLE-R~\cite{middler2005} show how middleware can provide clustering, routing, and replication control for replicated databases. 
More generally, replication middleware has long studied how to distribute requests across replicas while preserving consistency and availability~\cite{dbsurvey2000}. 
Practical systems such as Pgpool-II~\cite{pgpool} and HAProxy~\cite{haproxy} provide pooling, health checking, load balancing, and failover capabilities that are widely used in front of database services. 
FLARe~\cite{flare2007} is also relevant in showing that post-fault redirection should take load into account rather than treating failover as a purely availability-driven operation. 
Still, these systems primarily address replica management and request redirection. 
They do not formulate the control problem that is central in our setting: once the active set shrinks from \(K\) to \(K^+\), how should the remaining replicas be reorganized so that service remains stable and differentiated across priority levels, rather than simply routing all sessions to the available nodes in a homogeneous way?

A third line of work studies prioritization under resource pressure. 
Classic work on priority in DBMS scheduling~\cite{prioritydbms1989aa} considers how to allocate database resources across requests with different importance levels. 
More broadly, Dean and Barroso~\cite{tailatscale2013} show that tail latency can dominate user-visible performance under stress, making uniform treatment especially costly when capacity is reduced. 
CockroachDB~\cite{cockroachdb2024} therefore implements priority-aware admission control to protect critical work under contention. 
These works share the intuition that stress should not be handled uniformly. 
However, they operate either inside the DBMS or as general overload-control mechanisms. 
Our focus is different: we study how priority should shape failure handling at the replica-pool level after a fail-stop fault reduces available capacity.

In contrast, our paper studies priority-aware failure handling for fixed replica pools. 
The problem is not only to detect failures or continue routing around them, but also to define how the remaining pool should be reorganized after capacity loss so that high-priority service remains protected and low-priority impact stays measurable.

\section{Threats to Validity}\label{sec:threats}

\paragraph*{Failure model and recovery scope}
This paper studies fail-stop replica failures. 
A failed replica is treated as unavailable, excluded from new session assignments, and later returned to the active set. 
This abstraction isolates the post-fault regime targeted by PLB, but it does not cover other failure modes studied in distributed systems, including fail-slow behavior, intermittent reachability, partial network partitions, replica lag, or correlated failures~\cite{failslow2021,slowfault2025,alfatafta2020}. 
The current design also treats a rejoined replica as eligible for routing once it returns to the active set. 
It does not implement quarantine, staged reintroduction, or explicit cold-cache handling. 
Such cases require health and recovery mechanisms beyond the repair-to-target policy evaluated in this paper.

\paragraph*{Workload and deployment scope}
The evaluation uses a replicated PostgreSQL deployment and a TPC-H analytical workload with service-class-aware sessions. 
This setting matches the target environment of read-heavy replicated services under fixed capacity, but it does not cover OLTP or mixed read/write workloads, other DBMSs, other replication mechanisms, or heterogeneous deployments. 
In those settings, contention and recovery behavior may be affected by transaction conflicts, update propagation, replica freshness, or hardware imbalance. 
The empirical results should therefore be interpreted in the context of the evaluated environment. 
The comparisons are nevertheless made under the same workload, replica budget, and failure schedule, isolating the relative behavior of PLB with respect to static isolation and priority-agnostic sharing.

\paragraph*{Control granularity}
PLB operates at the middleware layer by updating replica roles and controlling subsequent session assignments. 
It does not migrate running sessions, preempt in-flight queries, or schedule individual queries inside the DBMS. 
Consequently, PLB can reorganize the replica pool for newly admitted sessions, but it cannot directly remove work already admitted before or during a failure. 
This limitation occurs during rejoin phases, where service-level recovery can lag behind replica availability. 
Extending PLB with backlog-aware admission, staged rejoin policies, or query-level mechanisms is left for future work.
\section{Conclusion}\label{sec:conclusion}
This paper studied priority-aware failure handling for replicated database services operating under a fixed replica budget. 
We focused on the post-fault regime following replica failures, during which the service must continue with fewer replicas and cannot assume immediate replacement capacity. 
In this setting, failure handling is not limited to excluding failed replicas from routing; it must also define how the remaining capacity is allocated across service classes.

We presented a repair-to-target failure-handling policy implemented in PLB that updates the role assignments of healthy replicas after faults and rejoins. 
PLB keeps the replica pool shared but reorganizes it based on the current active set, ensuring Premium traffic receives preferential access to capacity. 
This addresses two limitations of common deployment choices: static isolation can strand healthy capacity after asymmetric failures, whereas priority-agnostic sharing uses the remaining replicas without preserving service differentiation.

The evaluation shows that PLB improves both capacity use and service differentiation after failures. 
Under a Premium-side fault, PLB improves median Premium goodput retention by 26--28 percentage points and raises CPU utilization among available replicas by 18 percentage points in the Premium-heavy workload. 
Under cascading failures, PLB achieves more than \(2\times\) higher Premium goodput in the most severe phase and reduces Premium p95 latency by \(18.2\%\) relative to shared round-robin. 
The rejoin phases further show that replica recovery and service-level recovery need not coincide, since work accumulated while capacity was missing can affect later completions. 
This identifies recovery-tail handling as a separate control problem, with future work including staged replica reintroduction, backlog-aware admission, fail-slow handling, and broader workload classes.

\section*{Acknowledgment}
This work received funding from the France 2030 program, managed by the French
National Research Agency under grant agreement No.~ANR-23-PECL-0003.

\clearpage

\end{document}